\documentstyle[11pt,newpasp,twoside,epsf]{article}
\begin{document}
\title{Optical Counterparts to Damped Lyman Alpha Systems}
\author{Ariyeh H. Maller}
\affil{Physics Department, University of California, Santa Cruz, CA 95064}
\author{Jason X. Prochaska}
\affil{Observatories of the Carnegie Institution of Washington, Pasadena 
CA 91101}
\author{Rachel S. Somerville}
\affil{Racah Institute of Physics, The Hebrew University, Jerusalem, 
91904, Israel}
\author{Joel R. Primack}
\affil{Physics Department, University of California, Santa Cruz, CA 95064}

\begin{abstract}
Previously we have shown (Maller et al, 1998) that the kinematics of
Damped Lyman Alpha Systems (DLAS) as measured by Prochaska and Wolfe
(1998) can be reproduced in a multiple disk model (MDM) if the gaseous
disks are of sufficient radial extent.  Here we discuss this model's
predictions for the relationship between DLAS and Lyman break galaxies
(LBGs), which we here take to be objects at $z \sim 3$ brighter than
${\mathcal{R}}=25.5$. We expect that future observations of the correlations 
between DLAS and LBGs will provide a new data set able to discriminate 
between different theoretical models of the DLAS. 
Djorgovski (1997) has already detected a few optical
counterparts and more studies are underway.
\end{abstract}


\section{Summary}

We have used the Semi Analytic Models (SAMs) of Somerville \& Primack
(1999) to determine the distribution of galaxies in a dark matter
halo, and the amount of cold gas in each galaxy.  The SAMs also
contain the star formation history of each galaxy so we can explore
the optical properties of the galaxies in the halos that give rise to
DLAS.  Here we present results on the optical counterparts from the
two models discussed in Maller (1999), in which we matched the
kinematic data with thicker and less radially extended, or thinner,
more radially extended gas disks.  We show both models here only to
demonstrate that the optical properties are not highly sensitive to
the details of the gas modeling.  We only refer to optical
counterparts that reside in the same virialized halo that produces the
DLAS. The contribution of LBGs in neighboring dark matter halos will
be explored in future work, but is expected to be relatively
unimportant.

\begin{figure}
\plottwo{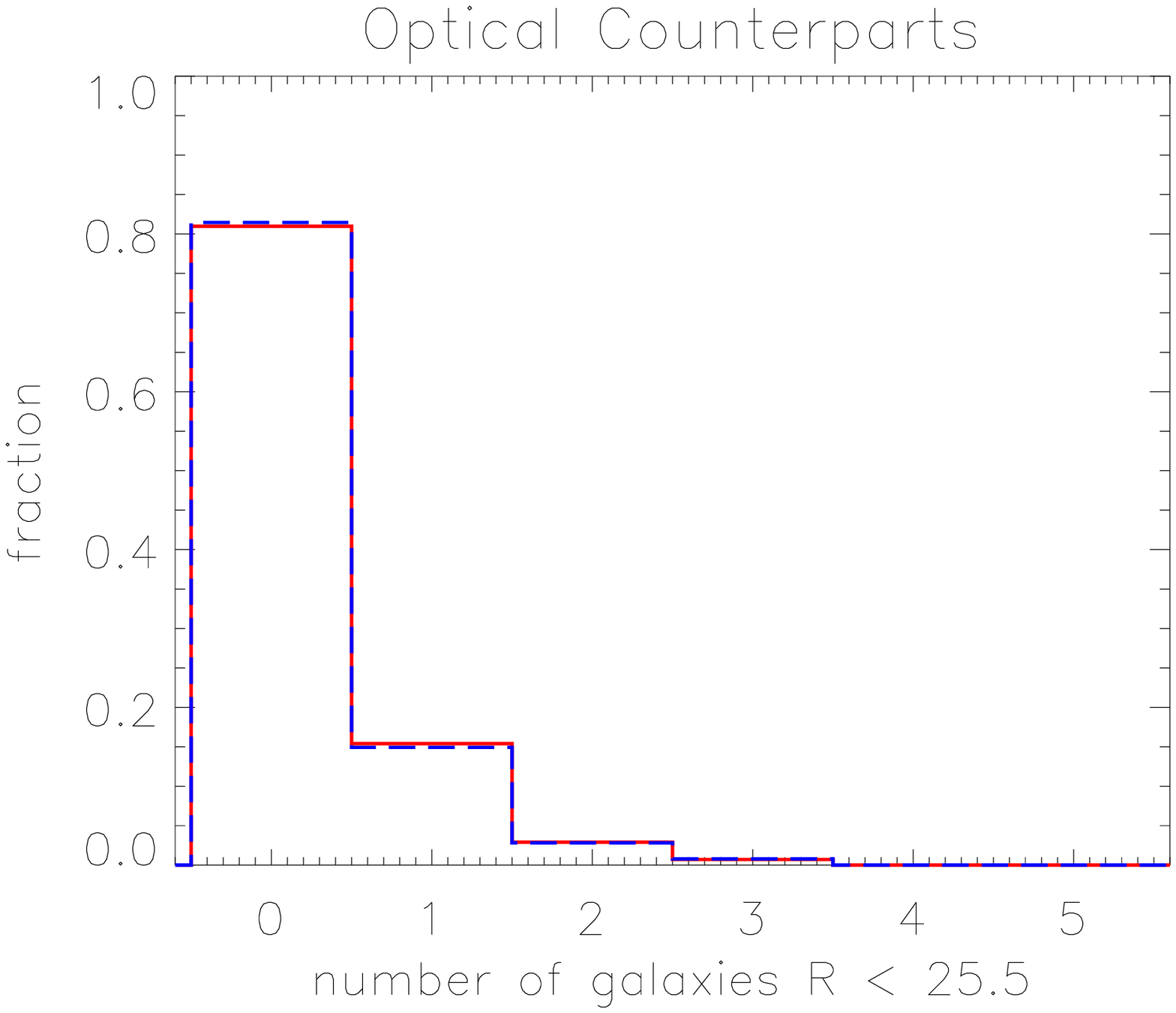}{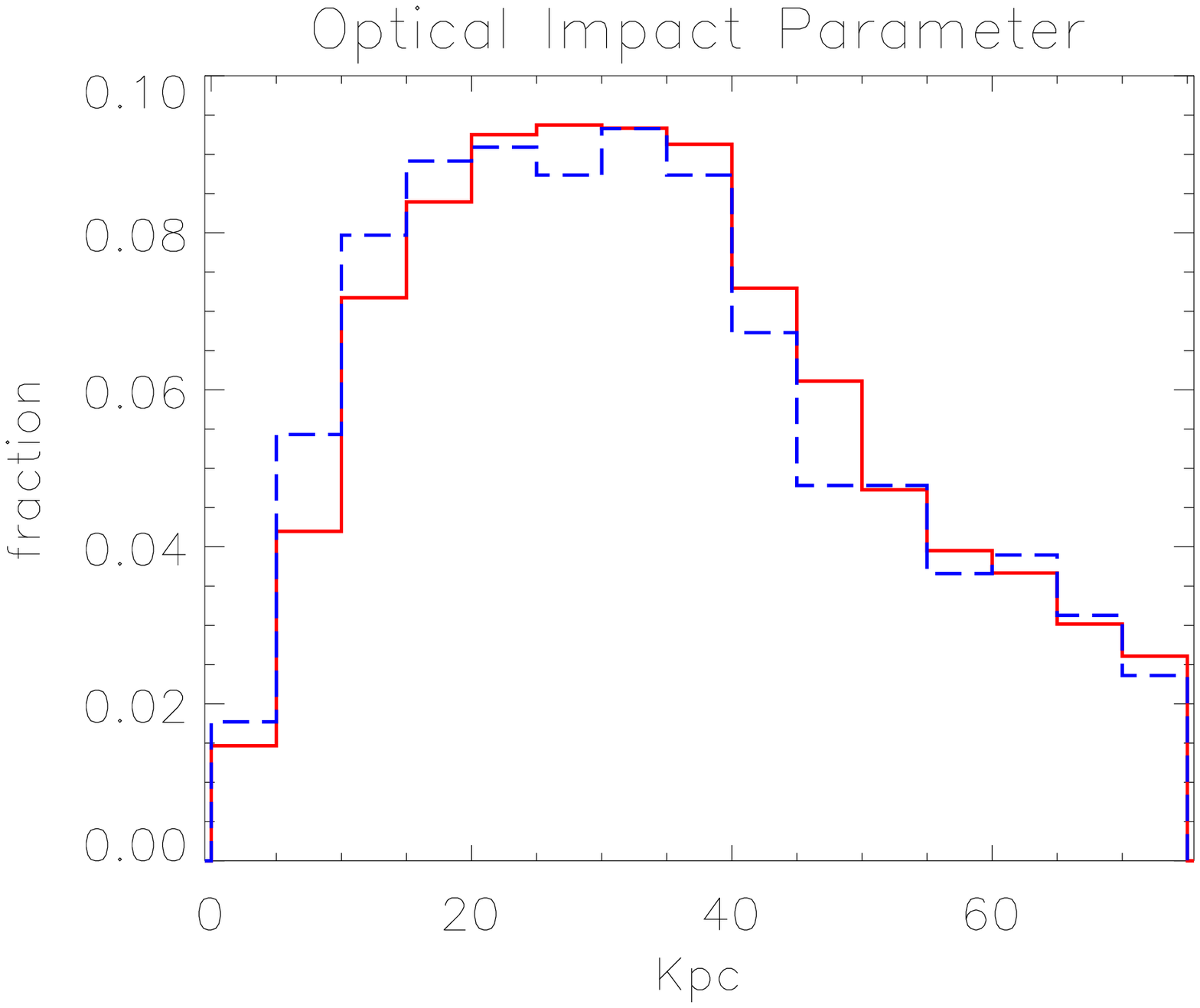}
\caption{ a) The distribution of the number of 
optical counterpart of DLAS.
b) The distribution of optical impact parameters.  In both panels the 
solid line is the thicker gas disk model, while the dashed line is for the
thinner gas disks.  
}\label{figopt}
\end{figure}

The properties of the optical counterparts will place strong
constraints on DLAS models.  One constraint is the number of DLAS with
optical counterparts.  Figure~\ref{figopt}a shows the distribution
that we see in our models.  Eighty percent of DLAS do not have an
optical counterpart with ${\mathcal{R}} < 25.5$, while a rare five
percent contain two or more such galaxies in the same halo.  Lastly we
show the distribution of optical impact parameter
(Figure~\ref{figopt}b) in our models.  The optical impact parameter is
the physical distance between the line of sight to the quasar and the
centroid of the light distribution of the LBG.  We obtain a broad
distribution of optical impact parameter values in our model.  Because
of the large radial extent of our gas disks, the DLAS are often many
stellar disk scale lengths from the center of the light distribution.
Also in the MDM scenario, with many galaxies in a single halo, sometimes the 
galaxy bright enough to be identified as an optical counterpart is not one of 
the galaxies giving rise to the DLAS: in this case very large separations are  
possible.  Thus we expect the
predictions about the optical impact parameter to be unique to the
multiple disk model, and a useful way of distinguishing it from other
models.

\end{document}